\documentclass[a4paper,11pt]{article}
\pdfoutput=1
\usepackage{jcappub,natbib}
\usepackage{amsmath,amssymb,color}

\title{\centering Wandering in the Lyman-alpha Forest: A Study of Dark Matter-Dark Radiation Interactions}

\author{Rebecca Krall,}
\author{Francis-Yan Cyr-Racine,}
\author{and Cora Dvorkin}
\affiliation{Harvard University, Department of Physics, \\Cambridge, MA 02138, USA}

\emailAdd{rkrall@physics.harvard.edu, dvorkin@physics.harvard.edu, fcyrraci@physics.harvard.edu}

\abstract{The amplitude of large-scale matter fluctuations inferred from the observed Sunyaev-Zeldovich (SZ) cluster mass function and from weak gravitational lensing studies, when taken at face value, is in tension with measurements of the cosmic microwave background (CMB) and baryon acoustic oscillation (BAO). In this work, we revisit whether this possible discrepancy can be attributed to new interactions in the dark matter sector. Focusing on a cosmological model where dark matter interacts with a dark radiation species until the epoch of matter-radiation equality, we find that measurements of the Lyman-alpha flux power spectrum from the Sloan Digital Sky Survey provide no support to the hypothesis that new dark matter interactions can resolve the possible tension between CMB and large-scale structure (LSS). Indeed, while the addition of dark matter-dark radiation interactions leads to an improvement of $2\Delta\ln\mathcal{L}=12$ with respect to the standard $\Lambda$ cold dark matter ($\Lambda$CDM) model when only CMB, BAO, and LSS data are considered, the inclusion of Lyman-alpha data reduces the improvement of the fit to $2\Delta\ln\mathcal{L}=6$ relative to $\Lambda$CDM. We thus conclude that the statistical evidence for new dark matter interactions (largely driven by the Planck SZ dataset) is marginal at best, and likely caused by systematics in the data. We also perform a Fisher forecast analysis for the reach of a future dataset composed of a CMB-S4 experiment combined with the Large Synoptic Survey Telescope galaxy survey. We find that the constraint on the effective number of fluid-like dark radiation species, $\Delta N_{\rm fluid}$, will be improved by an order of magnitude compared to current bounds.  
}


\arxivnumber{1705.08894}

\begin{document}
\maketitle
\section{Introduction}
Dark matter (DM) forms the gravitational backbone upon which baryonic matter accretes to form galaxies and clusters. The cold dark matter (CDM) paradigm \cite{Davis:1985rj,Blumenthal:1984bp,Blumenthal:1982mv,1981ApJ...250..423D} has so far been extremely successful at describing the large-scale distribution of galaxies and the structure of the anisotropies in the cosmic microwave background (CMB). Detailed observations of the CMB \cite{Aghanim:2015xee,Ade:2015xua} have provided us with an exquisite snapshot of the Universe as it stood about 380,000 years after the Big Bang. At that time, the data show that the Universe was mostly smooth and homogeneous except for very small density fluctuations. In our current understanding of structure formation, these small perturbations form the seeds which eventually evolve, through the influence of gravity, into all the rich structure we observe in the Universe today. If this scenario is correct, CMB measurements can be used to predict the properties of structure in the low-redshift universe. By comparing these predictions to the actual observations of the large-scale structure (LSS) of the Universe, we can thus test the consistency of the standard structure formation paradigm based on CDM. Such comparison is often phrased in terms of the quantity $\sigma_8$, which stands for the amplitude of matter fluctuations at scales of $8h^{-1}$ Mpc. 

Estimates of the value of $\sigma_8$ from recent LSS measurements based on the Sunyaev-Zeldovich (SZ) cluster mass function \cite{Ade:2013lmv,Ade:2015fva} and weak gravitational lensing \cite{Heymans:2013fya,2017MNRAS.467.3024L} appear in tension with $\Lambda$CDM predictions based on fits to CMB and baryon acoustic oscillation (BAO) data.  While the tension is likely caused by systematics in the data, it could also be the result of new physics related to DM. One approach to reconcile the CMB with the discrepant LSS measurements is to suppress the growth of structure by coupling DM to some form of dark radiation (DR) \cite{Goldberg:1986nk,HOLDOM198665,1992ApJ...398..407G,1992ApJ...398...43C,Boehm:2001hm,Boehm:2000gq,Foot:2003jt,Foot:2002iy,Foot:2004wz,Boehm:2004th,Foot:2004pa,Feng:2008mu,Ackerman:2008gi,Feng:2009mn,ArkaniHamed:2008qn,Kaplan:2009de,2010PhRvD..81h3522B,Kaplan:2011yj,Behbahani:2010xa,Das:2012aa,Hooper:2012cw,Aarssen:2012fx,Cline:2012is,Tulin:2013teo,Tulin:2012wi,Baldi:2012ua,Dvorkin:2013cea,Cyr-Racine:2013ab,Cline:2013zca,Chu:2014lja,Cline:2013pca,Cyr-Racine:2013fsa,Bringmann:2013vra,Archidiacono:2014nda,2014PhLB..739...62K,Choquette:2015mca,Chacko:2015noa,Buen-Abad:2015ova,2016PhRvD..93l3527C,Chacko:2016kgg,Kamada:2016qjo,Ko:2016fcd,Ko:2016uft,Ko:2017uyb,Foot:2016wvj,Foot:2014uba,Foot:2013vna,Foot:2011ve}. In several of these models, achieving a sufficient suppression of structure on scales probed by $\sigma_8$ tends to introduce further problem on smaller scales. However, a specific category \cite{Buen-Abad:2015ova,Lesgourgues:2015wza,Chacko:2016kgg,Ko:2016fcd,Ko:2016uft} of models where the interaction rate between DM and DR tracks the Hubble rate during the radiation-dominated era have the potential to address the tension without introducing new problems.  

The key ingredient of these models is the presence of a DM-DR scattering amplitude scaling as $|\mathcal{M}(q)|\propto 1/q^2$, where $q$ is the momentum transferred in a collision. For instance, Ref.~\cite{Buen-Abad:2015ova} realizes this by introducing a non-abelian massless gauge boson coupling to DM. In this scenario, the DR forms a tightly-coupled perfect fluid which provides a weak drag force on DM before matter-radiation equality, hence slowing the growth of structure on scales entering the causal horizon before that epoch. Ref.~\cite{Lesgourgues:2015wza} then used this non-abelian interacting DM model to reanalyze the apparently discrepant CMB and LSS data, finding that the model could alleviate the $\sigma_8$ tension between these datasets. Taken at face value, their analysis implies a statistically significant detection of DM-DR interaction, with an improvement of the $-2\Delta\ln\mathcal{L}$ statistics of 11.4 relative to $\Lambda$CDM. We note that this difference is mostly driven by data from the Planck SZ clusters \cite{Ade:2013lmv,Ade:2015fva}.

In this paper, we revisit the analysis performed in Ref.~\cite{Lesgourgues:2015wza} by adding Lyman-$\alpha$ forest flux power spectrum data from the Sloan Digital Sky Survey (SDSS) \cite{McDonald:2004eu,McDonald:2004xn}. We find that  Lyman-$\alpha$ data disfavor the low values of $\sigma_8$ preferred by the Planck SZ clusters, hence reducing the statistical significance of the evidence for DM-DR interaction.

A summary of the DM-DR interaction model is presented in Section \ref{sec:model}. In Section \ref{sec:analysis}, we describe in detail our Bayesian analysis of the CMB, BAO, LSS, and Lyman-$\alpha$ data in light of the DM-DR interaction model. The results from this analysis are presented in Section \ref{sec:results}. In Section \ref{sec:forecasts}, we perform a Fisher forecast to determine the projected constraints on the parameters of the DM-DR model from a combination of the Large Synoptic Survey Telescope (LSST) photometric survey and the next generation CMB-S4 experiment. We finally summarize our results and their implications in Section \ref{sec:discussion}.

\section{Dark matter interaction model}\label{sec:model}
For the type of model considered in this work, the standard $\Lambda$CDM scenario is extended by adding a massless DR component capable of scattering with the nonrelativistic DM at early times. As in the model proposed in Ref.~\cite{Buen-Abad:2015ova}, we treat the DR as a perfect fluid with no viscosity and a speed of sound $c_{\rm s}^2$ = 1/3. The DR energy density is parameterized as an effective number of neutrino species 
\begin{equation}
\Delta N_{\textrm{fluid}}=N_{\textrm{dr}}\left(\frac{T_{\textrm{dr}}}{T_\nu}\right)^4\times\left\{\begin{array}{ll}\frac{8}{7}\textrm{ (bosonic)}\\
1 \textrm{ (fermionic)},
\end{array} \right.
\end{equation}
where $T_{\rm dr}$ is the temperature of the DR, $T_\nu$ is the temperature of the Standard Model (SM) neutrinos, and $N_{\rm dr}$ is the total number of DR species. 

We focus here on models where the interaction between DM and DR is mediated by a massless particle, hence leading to a scattering amplitude scaling as $|\mathcal{M}(q)|\propto 1/q^2$, where $q$ is the momentum transfer. For such interactions, the linearized collision term between DM and DR can be computed as described in Ref.~\cite{2016PhRvD..93l3527C,Binder:2016pnr}, and results in a drag force on the DM by the DR quantified by the momentum-transfer rate $\Gamma$. Specifically, a non-relativistic DM particle with velocity $\vec{v}$ traveling through the thermal DR bath experiences an acceleration $\dot{\vec{v}}=-a\Gamma\vec{v}$, where the overdot corresponds to the derivative with respect to conformal time, and $a$ is the scale factor. For DM-DR interactions mediated by a massless particle, $\Gamma$ scales as $T_{\rm dr}^2$. Given the value of the momentum-transfer rate today ($\Gamma_0$), its value at another time is then simply given by
\begin{equation}\label{eq:gamma}
\Gamma=\Gamma_0\left(\frac{T}{T_0}\right)^2,
\end{equation}
where $T$ is the photon temperature, $T_0$ is the CMB temperature today, and the above scaling is valid as long as no entropy dump occurs in the dark sector. A key feature of such models is that the momentum-transfer rate has the same temperature dependence as the Hubble rate during radiation domination. This implies that $\Gamma$ tracks the Hubble expansion rate until the epoch of matter-radiation equality and DM kinetic decoupling is thus significantly delayed compared to the standard CDM scenario, leading to a suppression of structures on small scales. Furthermore, the self-interacting nature of the DR implies that its impact on the CMB and structure formation is significantly different than standard free-streaming neutrinos \cite{Bashinsky:2003tk,Hou:2011ec,2016JCAP...01..007B}.

As an example, Ref.~\cite{Buen-Abad:2015ova} achieves the $T^2$ scaling by having a non-Abelian DM component transforming in the fundamental representation of a dark $SU(N)$ gauge group. Associated with the $SU(N)$ dark symmetry are $N^2-1$ dark gluons forming the massless DR bath. In the early universe at temperatures above the DM mass, the dark gluons are taken to be in thermal equilibrium with the SM bath through their interactions with DM, which is taken to carry some electroweak charge. At temperatures on the order of the DM mass, DM freeze-out occurs, the dark gluons decouple from SM, and $T_{\rm dr}$ begins to evolve independently from the SM temperature. After DM freeze-out, the dark gluons form a self-interacting tightly-coupled fluid interacting with the DM.

The evolution of DM and DR perturbations in the Newtonian gauge is governed by the set of equations \cite{Ma:1995ey}
\begin{align}
\dot{\delta}_{\textrm{dm}}&=-\theta_{\textrm{dm}}+3\dot{\phi},\\
\dot{\theta}_{\textrm{dm}}&= -\frac{\dot{a}}{a}\theta_{\textrm{dm}}+a\Gamma(\theta_{\textrm{dr}}- \theta_{\textrm{dm}}) + k^2\psi,\\
\dot{\delta}_{\textrm{dr}}&=-\frac{4}{3}\theta_{\textrm{dr}}+4\dot{\phi},\\
\dot{\theta}_{\textrm{dr}}&= k^2\frac{\delta_{\textrm{dr}}}{4}+\frac{3}{4}\frac{\rho_{\textrm{dm}}}{\rho_{\textrm{dr}}}a\Gamma(\theta_{\textrm{dm}}-\theta_{\textrm{dr}}) + k^2\psi,
\end{align}
where $k$ is the Fourier wavenumber, $\Gamma$ is defined in Eq.~\eqref{eq:gamma}, and where $\delta_{ \textrm{dm} }$ ($\delta_{ \textrm{dr} }$) and $\theta_{ \textrm{dm} }$ ($\theta_{ \textrm{dr} }$) are the DM (DR) density and velocity divergence perturbations, respectively. The average energy densities of DM and DR are $\rho_\textrm{dm}$ and $\rho_\textrm{dr}$, and the scalar metric perturbations in the conformal Newtonian gauge are $\phi$ and $\psi$. We note that the self-interacting nature of the DR ensures that the higher multipoles of the DR Boltzmann hierarchy remain negligible throughout the history of the Universe. Relative to the $\Lambda$CDM paradigm, the DM-DR interaction models considered here have two additional parameters: $\Gamma_0$ and $\Delta N_{\rm fluid}$. We now perform a quantitative analysis of such models. 

\section{Analysis}\label{sec:analysis}

We implemented the DM-DR model in the Boltzmann code CLASS \cite{Lesgourgues:2011re,Blas:2011rf}, and we performed several Markov Chain Monte Carlo (MCMC) likelihood analyses for our 8-parameter DM-DR interaction model, as well as for the standard $\Lambda$CDM scenario. We utilize the same data sets used in Ref.~\cite{Lesgourgues:2015wza}, namely: CMB, BAO and LSS (see below for details). In addition to those data sets, we also include for the first time in this context constraints on the matter power spectrum from the Lyman-$\alpha$ forest flux power spectrum measurements from the SDSS. More specifically, we use the following cosmological data sets:
\begin{itemize}
\item \textbf{CMB:} Planck 2015 temperature + low-$\ell$ polarization data \cite{Aghanim:2015xee}.
\item \textbf{BAO:} measurements of the acoustic-scale distance ratio $D_V /r_{\textrm{drag}}$ at $z = 0.106$ by 6dFGS \cite{2011MNRAS.416.3017B}, at $z = 0.1$5 by SDSSMGS \cite{Ross:2014qpa}, at $z = 0.32$ by BOSS- LOWZ \cite{Anderson:2013zyy}, and anisotropic BAO measurements at $z = 0.57$ by BOSS-CMASS-DR11 \cite{Anderson:2013zyy}.
\item \textbf{LSS:} Planck 2015 lensing likelihood \cite{Ade:2015zua}, the constraint $\sigma_8(\Omega_{\rm{m}}/0.27)^{0.46}= 0.774\pm0.040$ (68\% CL) from the weak lensing survey CFHTLenS \cite{Heymans:2013fya}, and the constraint $\sigma_8(\Omega_{\rm{m}}/0.27)^{0.30}= 0.782\pm0.010$ (68\% CL) from the Planck SZ cluster mass function \cite{Ade:2013lmv}.
\item \textbf{Ly$\alpha$:}  measurements of the Lyman-$\alpha$ flux power spectrum from the SDSS \cite{McDonald:2004eu,McDonald:2004xn}.
\end{itemize}

For our likelihood evaluation, we modified the publicly available code \textsc{MontePython} \cite{Audren:2012wb} to include the Lyman-$\alpha$ forest likelihood from the SDSS. We did several MCMC runs for four different combinations of these datasets, for both $\Lambda$CDM and for the DM-DR interaction model. These combinations are: CMB+BAO, CMB+BAO+LSS, CMB+BAO+Ly$\alpha$, and CMB+BAO+LSS+ Ly$\alpha$. 

The parameters of our model are $\left\lbrace \Omega_{\rm b}h^2, \Omega_{\rm dm}h^2, \theta, \ln{(10^{10}A_{\rm s})}, n_{\rm s}, \tau, \Delta N_{\textrm{fluid}}, {10^7\Gamma_0}\right\rbrace$, where $\Omega_{\rm b}h^2$ and $\Omega_{\rm dm}h^2$ are the physical baryon and DM densities, respectively, $\theta$ is the angular size of the horizon at recombination, $A_{\rm s}$ is the amplitude  of the initial curvature power spectrum at $k=0.05$ Mpc$^{-1}$, $n_{\rm s}$ is the spectral index, $\tau$ is the optical depth to reionization,  $\Delta N_{\textrm{fluid}}$ is the effective number of DR species, and $\Gamma_0$ is the current value of the momentum-transfer rate between the DM and the DR.  We set two massless and one massive neutrino species with a total mass of $0.06$ eV, and with an effective neutrino number of $N_{\rm{eff}} = 3.046$. The primordial helium abundance is inferred from standard Big Bang Nucleosynthesis, as a function of $\Omega_{\rm b}$, following Ref. \citep{Hamann:2007sb}. 

All of these parameters have flat priors.  For the interaction rate we set the lower bound on the prior to $\Gamma_0 \geq 0$, while for the optical depth to reionization, we set the lower bound of the prior to $\tau\geq 0.04$. We also impose the prior $\Delta N_{\textrm{fluid}} \geq 0.07$ since this is the smallest allowed value if the dark sector decoupled from the SM near the weak scale. While there are ways around this limit (for instance if the dark sector is never in thermal equilibrium with the SM), we focus on the broad class of models respecting this bound. As an example, the non-Abelian DM-DR model \cite{Buen-Abad:2015ova} predicts discrete values for $\Delta N_{\rm{fluid}}=0.07(N^2-1)$, while a model where DM couples to a massless dark photon that also interacts with $N_{\rm f}$ massless fermions predicts $\Delta N_{\rm{fluid}}=0.07(1+\frac{7}{4}N_{\rm f})$.
\section{Results}\label{sec:results}
\begin{figure}[h!]
\begin{center}
\includegraphics[width=0.85\textwidth]{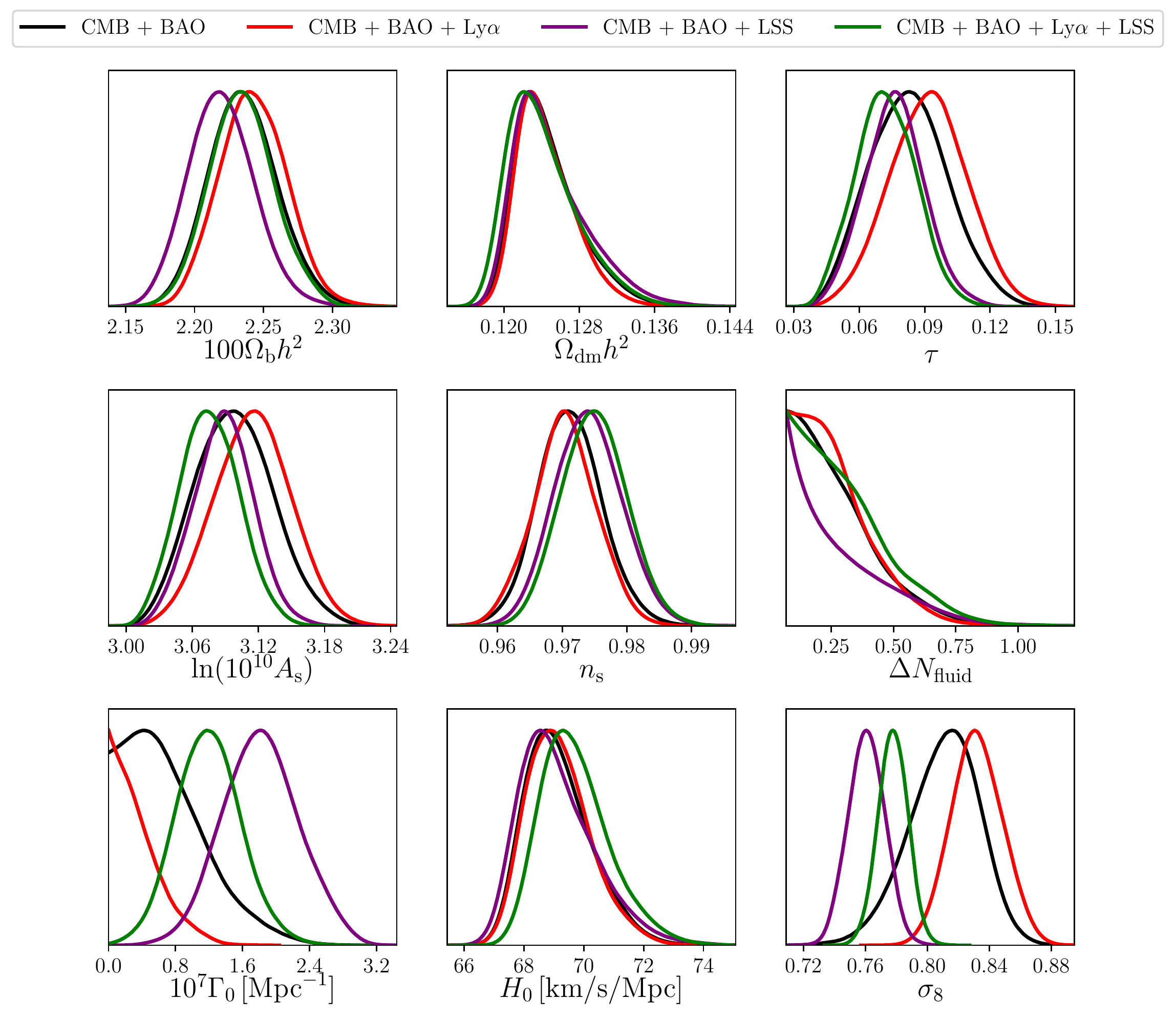}
\caption{Marginalized posterior probability distributions for the most relevant parameters of the DM-DR interaction model for CMB and BAO data (black), combined with Lyman-$\alpha$ (red), LSS (purple), and Lyman-$\alpha$ and LSS (green).}\label{fig:1d}
\end{center}
\end{figure}

The parameter confidence regions from our analysis are listed in Table \ref{table:results}. The bottom row gives the $-2\Delta\ln\mathcal{L}$ values between the DM-DR interaction model and the standard 6-parameter $\Lambda$CDM paradigm. We also display in Figure \ref{fig:1d} the marginalized posterior probability distributions for the parameter set $\{\Omega_{\rm b}h^2, \Omega_{\rm dm}h^2, \tau, \ln(10^{10} A_{\rm s}), n_{\rm s}, H_0, \sigma_8, \Delta N_{\textrm{fluid}},10^7\Gamma_0\}$. We observe that the constraints on the dark matter and baryon densities, the Hubble constant, the amplitude and tilt of the primordial spectrum of fluctuations, and on the optical depth to reionization are fairly robust from one data sets to the next.  As in Ref.~\cite{Lesgourgues:2015wza}, we find that the inclusion of LSS data favors a nonvanishing value of $\Gamma_0$, with an improvement of the log likelihood given by $-2\Delta\ln\mathcal{L}\simeq -12$. With the inclusion of Lyman-$\alpha$ forest data, we find that the DM-DR model still improves the fit relative to $\Lambda$CDM, but now the improvement becomes marginal with $-2\Delta\ln\mathcal{L}\simeq-5.8$. In addition, the Lyman-$\alpha$ data pull the preferred value of $\Gamma_0$ down by about $\sim$35\%, indicating that this latter data set does not favor the kind of matter power spectrum suppression (and corresponding small value of $\sigma_8$) necessary to fit the Planck SZ cluster data. 

\begin{table}[t!] 
\begin{center}
  \begin{tabular}{ | c | c | c | c | c |}
    \hline
    Parameter & CMB+BAO & CMB+BAO & CMB+BAO & CMB+BAO \\
    &&+LSS&+Ly$\alpha$&+LSS+Ly$\alpha$\\ \hline \hline
    $100\Omega_{\rm b}h^2$ &  $2.235^{+0.023}_{-0.026}$& $2.219^{+0.023}_{-0.025} $ & $2.243\pm 0.024$ & $2.234\pm 0.024$\\ 
    $\Omega_{\textrm{dm}}h^2$ & $0.1246^{+0.0021}_{-0.0040}$ & $0.1249^{+0.0023}_{-0.0047}$ & $0.1244^{+0.0021}_{-0.0036}$ & $0.1241^{+0.0024}_{-0.0045}$ \\
    $\Delta N_{\textrm{fluid}}$ & $<0.33$ (68\%) & $<0.34$ (68\%) & $<0.31$ (68\%) & $<0.37$ (68\%)\\
    $\Delta N_{\textrm{fluid}}$& $<0.60$ (95\%) & $<0.67$ (95\%) &$ <0.55$ (95\%) & $<0.65$ (95\%)\\
    $10^7\Gamma_0$ [\textrm{Mpc}$^{-1}]$ & $<1.64$ (95\%) & $1.82\pm 0.46$ & $<0.94$ (95\%)& $1.19\pm 0.39$\\
    $H_0$ [km/s/Mpc] & $69.15^{+0.82}_{-1.3} $ & $69.11^{+0.86}_{-1.5} $ & $69.16^{+0.84}_{-1.2}$ & $69.72^{+0.86}_{-1.3}$\\
    $\ln(10^{10}A_{\rm s})$ & $3.098^{+0.034}_{-0.038} $ & $3.088\pm 0.027$ & $3.115\pm 0.035$ & $3.076\pm 0.027$\\
    $n_{\rm s}$ &  $0.9712\pm 0.0050$ & $0.9741^{+0.0050}_{-0.0056}$ & $0.9704\pm 0.0049$ & $0.9750\pm 0.0051$\\
    $\tau$ & $0.083^{+0.017}_{-0.019}$ & $0.076\pm 0.014$ & $0.092\pm 0.018$ & $0.072\pm 0.014$\\
    $\sigma_8$ & $0.811^{+0.025}_{-0.020}$ & $0.761\pm 0.012$ & $0.832\pm 0.017$ &$0.7778\pm 0.0097$\\
    \hline
    $-2\Delta\ln\mathcal{L}/\Lambda$CDM & 1.5 & -11.98 & -0.22 & -5.84\\
    \hline
  \end{tabular}
  \caption{Mean values and confidence intervals for parameters of DM-DR interaction model. Unless otherwise noted, we display the 68\% confidence interval. The bottom row in the table lists the improvement in $-2\Delta\ln\mathcal{L}$ relative to $\Lambda$CDM.}
\label{table:results}
\end{center}
\end{table}

\begin{figure}[t!]
\begin{center}
\includegraphics[width=0.8\textwidth]{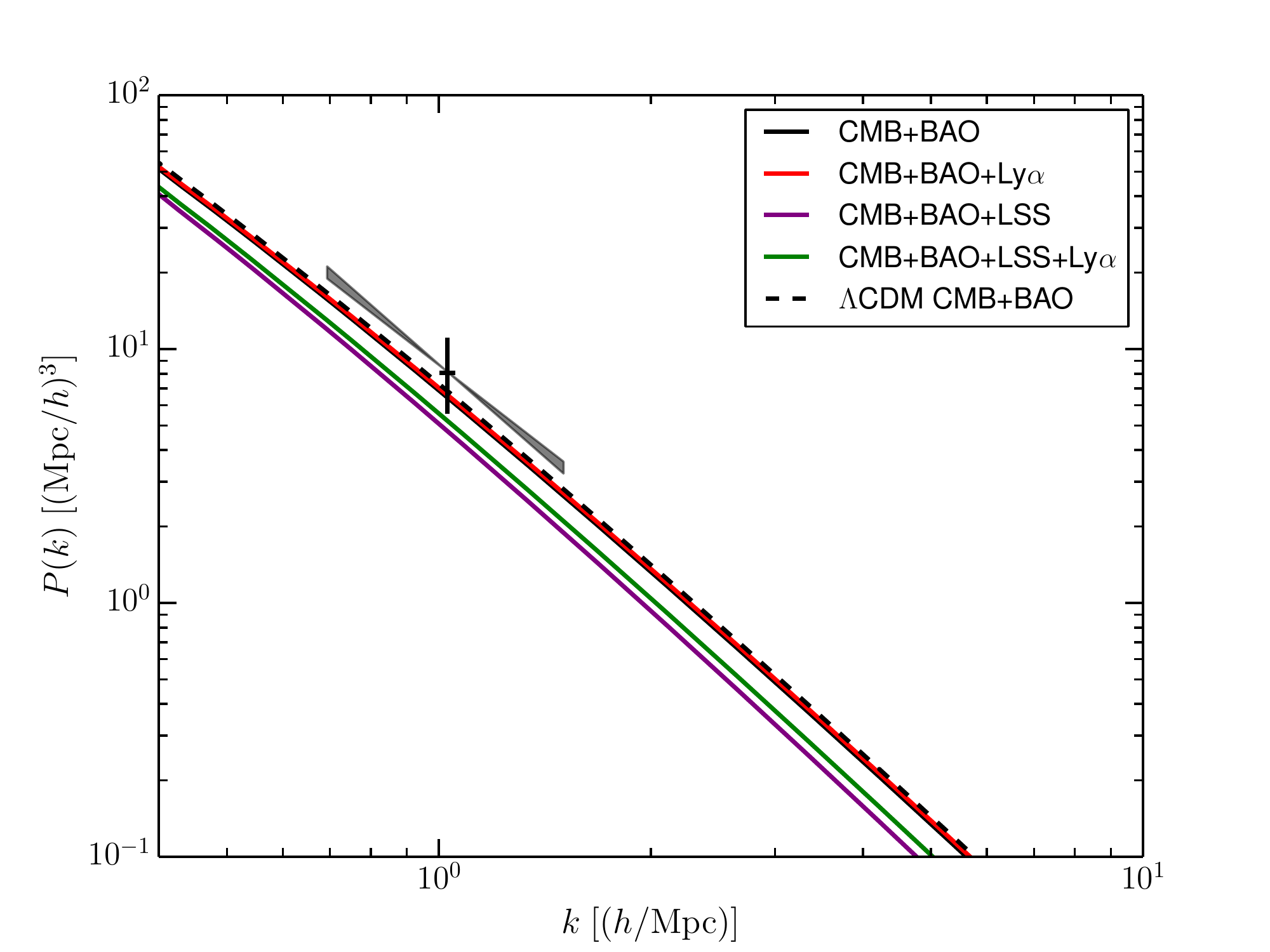}
\caption{The matter power spectra for the best-fit parameters for each dataset combination at $z=3$. The data point corresponds to the best-fit amplitude using Lyman-$\alpha$ data from Ref.~\cite{McDonald:2004xn}. The gray band shows the range of linear
matter power spectra slopes at $k = 1.03$ h/Mpc that are allowed at the 95\% CL limit. The error bar corresponds to the 95\% confidence region on the amplitude. All the solid lines show the power spectrum for the DM-DR interaction model, while the dashed line illustrate a $\Lambda$CDM model.}\label{fig:mpk}
\end{center}
\end{figure}

We illustrate the effect of the Lyman-$\alpha$ data in Figure \ref{fig:mpk}, where we display the matter power spectra for the best-fit parameters for each dataset combination at $z=3$. This figure indeed shows that the matter power spectra for the best-fit parameters of the dataset combinations that include LSS data lie outside of the Lyman-$\alpha$ $95$\% confidence interval (indicated with the black error bar). The addition of the Lyman-$\alpha$ data does raise the amplitude of the matter power spectrum, but it still falls significantly below the value preferred by the CMB+BAO+Ly$\alpha$ data combination. Clearly, a significant reduction of the Lyman-$\alpha$ error bar could potentially rule out the DM-DR interaction model as a solution to the discrepancy between Planck SZ data and the CMB. We estimate that the measurement error must be reduced by $\sim$60\% in order to exclude the DM-DR model at 3-$ \sigma$.\footnote{This improvement is determined by shrinking the error bars  of the Lyman-$\alpha$ measurement at $k=1.03\, h/\rm{Mpc}$, while keeping the mean value fixed, until P(k$=1.03\,h/\rm{Mpc}$) for the best-fit parameters with CMB+BAO+LSS+Ly$\alpha$ data lies 3-$\sigma$ away from the mean.} We further discuss possible improvement of the Lyman-$\alpha$ constraints in Section \ref{sec:discussion}.

It is instructive to look at how the different data likelihoods change when the DM-DR interactions are introduced. We display in Table \ref{table:likelihood} the likelihood decomposition for the best-fit model for the dataset combinations of CMB+BAO+ LSS and CMB+BAO+LSS+Ly$\alpha$, with and without the DM-DR interaction. The values in the table show that a significant portion of the likelihood improvement when including the DM-DR interaction (relative to $\Lambda$CDM) comes from the Planck SZ dataset, with minor contributions coming from Planck lensing and CFHTLenS. The Planck SZ data impose a tight constraint on $\sigma_8$ of $\sigma_8(\Omega_{\rm{m}}/0.27)^{0.30}= 0.782\pm0.010$, which is in the direction of lower $\sigma_8$ values. However, we also observe that introducing the DM-DR interactions also worsens the fit to the Lyman-$\alpha$ data. 

\begin{table}[t!]
\begin{center}
\begin{tabular}{| c | c | c | c | c |}
\hline
Likelihood & CMB+BAO & CMB+BAO & CMB+BAO & CMB+BAO \\
& +LSS & +LSS& +LSS+Ly$\alpha$  & +LSS+Ly$\alpha$ \\
&  ($\Lambda$CDM) & (DM-DR)  & ($\Lambda$CDM) &  (DM-DR) \\
\hline
\hline
Planck CMB & 5635.5 & 5634.8 & 5634.2 & 5634.8\\
BAO & 3.4 & 2.2 & 2.4 & 2.3\\
Ly$\alpha$ & - & - & 97.3 & 100.1\\
Planck SZ & 3.9 & 1.1 & 6.4 & 1.7\\
CFHTLenS & 0.7 & 0.5 & 1.1 & 0.5\\
Planck lensing & 7.3 & 6.2 & 6.7 & 5.6\\
\hline
Total -ln$(\mathcal{L})$ & 5650.8 & 5644.8 & 5748.1 & 5745.0\\
\hline
\end{tabular}
\caption{Likelihood decomposition for the best-fit $\Lambda$CDM and DM-DR models for two different combination of data sets.}
\label{table:likelihood}
\end{center}
\end{table}

\begin{figure}[t]
\begin{center}
\includegraphics[width=\textwidth]{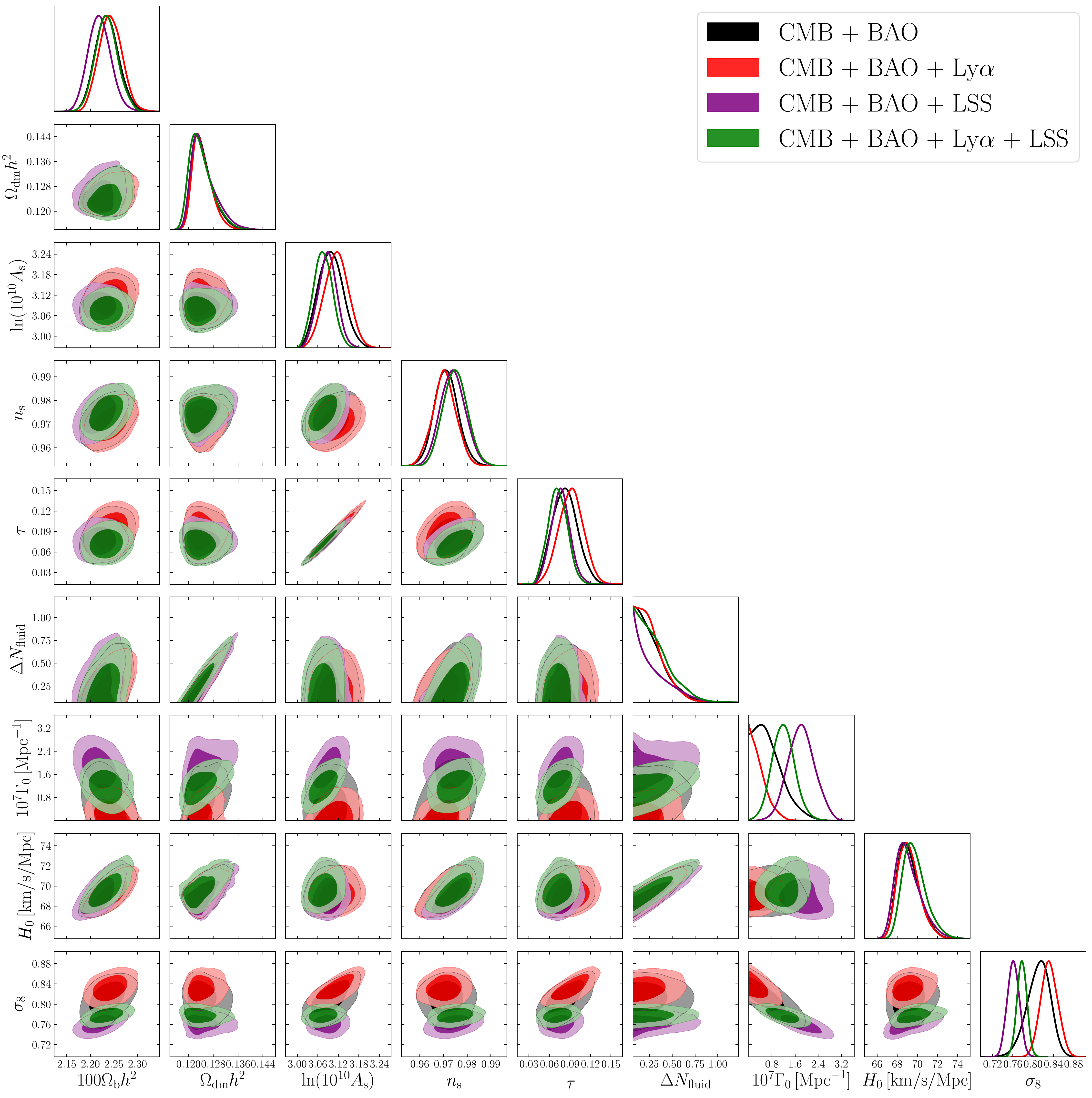}
\caption{The joint marginalized probability contours for the DM-DR model, for CMB and BAO data (black), combined with Lyman-$\alpha$ (red), LSS (purple), and Lyman-$\alpha$ and LSS (green). We display the 68\% and 95\% confidence regions.}\label{fig:2d}
\end{center}
\end{figure}

In Figure \ref{fig:2d}, we show the joint marginalized probability contours for the six standard cosmological parameters, in addition to $\Delta N_{\rm{fluid}}$, $\Gamma_0$, and $\sigma_8$. Most notable are the correlations between $H_0$ and $\Delta N_{\rm{fluid}}$, $\sigma_8$ and $\Gamma_0$, and $\Omega_{\rm dm}h^2$ and $\Delta N_{\rm{fluid}}$. The correlation between $H_0$ and $\Delta N_{\rm{fluid}}$ arises from the fact that an increase in radiation density contributes to the expansion rate of the universe. For $\sigma_8$ and $\Gamma_0$, the DM-DR interaction damps the matter power spectrum, and thus an increase in $\Gamma_0$ is strongly correlated to a decrease in $\sigma_8$. Since the epoch of matter-radiation equality is well-determined by CMB data, an increase of $\Delta N_{\rm{fluid}}$ must be compensated by a corresponding increase in $\Omega_{\rm dm}h^2$.

\section{Forecasts}\label{sec:forecasts}
Given that the current cosmological data sets do not have enough statistical power at the relevant scales to rule out with high significance the current best fit cosmology with DM-DR interactions favored by LSS data, we investigate how well future surveys will be able to test this hypothesis. In this section, we use the galaxy clustering as a tracer of matter fluctuations at intermediate scales between those probed by the CMB and the Lyman-$\alpha$ data, and we forecast the constraints on the DM-DR model from the photometric redshift survey expected from the Large Synoptic Survey Telescope (LSST) \cite{Abell:2009aa}, combined with expected observations of the CMB coming from the proposed CMB-S4 next generation CMB experiment \citep{Abazajian:2016yjj}. 

In our analysis, constraints from the CMB are taken into account by adding to the Fisher matrix for LSST a Fisher matrix for CMB-S4. We take as the fiducial model the mean values for the six standard cosmological parameters and the 95\% CL values for $\Gamma_0$ and $\Delta N_{\textrm{fluid}}$ from our MCMC analysis above, with CMB+BAO+LSS+Ly$\alpha$ data (see Table \ref{table:results} above). To forecast the parameter errors from LSST, we use the specifications in Ref.~\cite{Chen:2016vvw}. Specifically, the Fisher matrix for a galaxy survey is given by \cite{Kaiser:1987qv,Peacock1992}
\begin{equation}\label{eq:fisher_gal_clustering}
F_{ij}=\int_{-1}^1\int_{k_{\rm{min}}}^{k_{\rm{max}}}\frac{\partial\ln P_g(\bf{k})}{\partial\theta_i}\frac{\partial\ln P_g(\bf{k})}{\partial\theta_j}V_{\rm{eff}}(k,\mu)\frac{k^2dkd\mu}{2(2\pi)^2},
\end{equation}
where the $\theta_i$ are the parameters of the model, and $P_g$ is the redshift-space galaxy power spectrum, which can be determined from the matter power spectrum through
\begin{equation}\label{eq:galaxyps}
P_g(k,\mu)=[1+\beta\mu^2]^2b^2\hat{P}(k)e^{-c^2\sigma_z^2k^2\mu^2/H^2}.
\end{equation}
Here $b$ denotes the linear galaxy bias, $\beta = f/b$ with $f$ being the growth function (which we approximate as $\Omega_{\rm m}^{0.56}$ \citep{Chen:2016vvw}),  $\hat{P}$ is the smoothed matter power spectrum, $H$ is the Hubble rate, and $\mu$ is the cosine of the angle of the wavevector with respect to the line of sight. Additionally, $\sigma_z$ takes into account the accuracy in the redshifts $\sigma_{0\gamma z}$ and the intrinsic galaxy velocity dispersion $\sigma_{0v}$, and it is given by:
\begin{equation}
\sigma_z^2=(1+z)^2[\sigma_{0v}^2+\sigma_{0\gamma z}^2].
\end{equation}
For this analysis we take $\sigma_{0v}=400 \rm{km/s}/c$, and $\sigma_{0\gamma z}=0.04$, as in Ref.~\cite{Chen:2016vvw}.

The matter power spectrum in Eq.~\eqref{eq:galaxyps} is smoothed with a Gaussian window function
\begin{equation}
\hat{P}({\bf k})=\int d^3 {\bf k} P({\bf k'})|W(|{\bf k}-{\bf k'}|)|^2,
\end{equation}
where the window function is given by
\begin{equation}
|W(k)|^2=\frac{1}{(2\pi\sigma_W^2)^{3/2}}\exp\left(-\frac{k^2}{2\sigma_W^2}\right),
\end{equation}
and the width of the window is
\begin{equation}
\sigma_W=\frac{\sqrt{2\ln2}}{2\pi}k_{\rm{min}}.
\end{equation}

The value of $k_{\rm{min}}$ for each redshift bin is $2\pi(3V/4\pi)^{-1/3}$, with $V$ the volume of the survey.

We use the seven redshift bins, and their corresponding $k_{\rm{min}}$, $k_{\rm{max}}$, and number density of galaxies as defined in Table 2 of Ref.~\cite{Chen:2016vvw}. For each redshift bin, the bias used are: $\{1.053, 1.125, 1.126, 1.243, 1.243, 1.292, 1.497, 1.491\}$, which are the bias used for LSST in Ref.~\citep{Chen:2016vvw}. The effective survey volume is 
\begin{equation}\label{eq:surveyvolume}
V_{\rm{eff}}(k,\mu) = \int\left[\frac{n({\bf r})P_g(k,\mu)}{n({\bf r})P_g(k,\mu)+1}\right]^2d^3r\simeq \left[\frac{\bar{n}P_g(k,\mu)}{\bar{n}P_g(k,\mu)+1}\right]^2 V,
\end{equation}
in which $\bar{n}$ is the mean number density of galaxies. The volume for a survey over a fraction of the sky $f_{\rm sky}$ is
\begin{equation}
V=\frac{4\pi}{3}\times f_{\rm{sky}}[d_c(z_{\rm{max}})^3-d_c(z_{\rm{min}})^3], \label{eq:volume}
\end{equation}
where $d_c(z)$ is the comoving distance.
We take $\bar{n}$=\{0.154, 0.104, 0.064, 0.036, 0.017, 0.007, 0.002\} $[h^3\mathrm{Mpc}^{-3}]$, in each of the redshift bins. For each redshift bin, the volume $V$ is computed with Eq.~\eqref{eq:volume}, with $f_{\rm{sky}}$=0.58, and 
\begin{equation}
d_c(z) = \int_0^z\frac{c}{H(z')}dz'.
\end{equation}

We combine the above galaxy clustering forecast with future constraints expected from the proposed CMB-S4 experiment. In this case, the Fisher matrix takes the form
\begin{equation}\label{eq:fisher_CMB}
F_{ij}=\sum_l \frac{\partial \vec{C}^{T}_l}{\partial\theta_i}{\bf C}_l^{-1}\frac{\partial \vec{C}_l}{\partial\theta_j},
\end{equation}
where $\vec{C}_l = \{ C_l^{TT}, C_l^{EE}, C_l^{TE} \}$, and the elements of the covariance matrix ${\bf C}$ are given by:
\begin{equation}
{\bf C}(C_l^{\alpha\beta},C_l^{\gamma\delta})=\frac{1}{(2l+1)f_{\rm sky}}\left[(C_l^{\alpha\gamma}+N_l^{\alpha\gamma})(C_l^{\beta\delta}+N_l^{\beta\delta})+(C_l^{\alpha\delta}+N_l^{\alpha\delta})(C_l^{\beta\gamma}+N_l^{\beta\gamma})\right],
\end{equation}
where $\alpha,\beta,\gamma,\delta$ are $T,E$, and $f_{\rm sky}$ is the fractional area of sky used. We model the detector noise as: 
\begin{equation}
N_l^{\alpha\beta}=\delta_{\alpha\beta}\Delta^2_\alpha \exp\left(\frac{l(l+1)\theta^2_{\rm FWHM}}{8\ln 2}\right),
\end{equation}
where $\Delta_\alpha$ is the map sensitivity in $\mu$K-arcmin, $\theta_{\rm FWHM}$ is the beam width, and $\delta_{\alpha\beta}$ is the Kronecker delta.

For CMB-S4, we use $f_{\rm sky}=0.4$, $\Delta_{T}=1 \mu$K-arcmin, $\Delta_{E}=1.4 \mu$K-arcmin, and $\theta_{\rm FWHM}=3'$  \cite{Abazajian:2016yjj}. We consider $l\geq 30$, and set $l_{\rm max}=3000$. The $\theta_i$ in our Fisher matrix are the cosmological parameters: $\{\Omega_{\rm b}h^2$, $\Omega_{\rm dm} h^2$, $\Delta N_{\rm{fluid}}$, $10^7\Gamma_0$, $H_0$, $\ln(10^{10}A_{\rm s})$, $n_{\rm s}\}$. When considering LSST, we add to the previous array of parameters, the nuisance parameters $\sigma_{0v}$ and a bias $b_i$ for each bin. 

Table \ref{tab:fisher} displays the expected improvement from the proposed CMB-S4 experiment in combination with the future LSST survey for the DM-DR cosmological model. As can be seen there, we obtain an error on $\Delta N_{\rm fluid}$, for the combination of LSST and CMB-S4 of $\sigma(\Delta N_{\rm fluid})=0.011$, and for the $\Gamma_0$ coefficient is $\sigma(10^7\Gamma_0)=0.069$ Mpc$^{-1}$. 
\begin{table}
\begin{center}
\begin{tabular}{ c | c | c | c }
\hline
Parameter & fiducial &  CMB-S4 & CMB-S4 + LSST\\
\hline
\hline
100$\Omega_{\rm b}h^2$ & 2.2380& $\pm 0.0050$  & $\pm$0.0032\\
$\Omega_{\textrm{dm}}h^2$ & 0.12400 &$\pm 0.00134$&$\pm$0.00034\\
$\Delta N_{\textrm{fluid}}$ & 0.079 &$\pm 0.053$& $\pm$0.011\\
$10^7\Gamma_0$ [Mpc$^{-1}]$ & 1.148 &$\pm 0.323 $& $\pm$0.069\\
$H_0$ [km/s/Mpc]& 69.79 &$\pm 0.48$& $\pm$0.20\\
$\ln(10^{10}A_{\rm s})$ & 3.0750 & $\pm 0.0161$& $\pm$0.0093\\
$n_{\rm s}$ & 0.9754 &$\pm 0.0036$& $\pm$0.0022\\
\hline
\end{tabular}
\caption{Forecasted DM-DR 68\% parameter constraints for CMB-S4 and a LSST-like photometric survey. 
}\label{tab:fisher}
\end{center}
\end{table}
%

The forecasted constraints on the model parameters show that a combination of CMB-S4 and LSST data could provide a bound on $10^7 \Gamma_0$ that is about a factor of $\sim6$ times better than the constraints from our MCMC analysis using current data. This implies that a value of $10^7 \Gamma_0\sim1$ Mpc$^{-1}$ could be detected with high significance if this model was indeed describing the universe we live in. We caution, however, that the forecasted value of $\sigma(10^7\Gamma_0)$ depends quite strongly on the fiducial value of $\Gamma_0$ used in the analysis, and that it might be difficult to exclude much smaller values of $\Gamma_0$. We also obtain that the constraint on $\Delta N_{\rm fluid}$ will be improved by nearly an order of magnitude compared to current bounds (see Refs.~\cite{2016JCAP...01..007B,Brust:2017nmv}), hence severely restricting the presence of fluid-like DR in the early Universe. 

\section{Discussion}\label{sec:discussion}
In light of previous results \cite{Lesgourgues:2015wza}, we have re-examined the evidence for DM-DR  interaction within current cosmological data. We find that when adding to the CMB, BAO and LSS data another tracer of the matter fluctuations (i.e., the Lyman-$\alpha$ flux power spectrum measurements from the SDSS), the significance for the DM-DR model decreases from $-2\Delta\ln\mathcal{L}\simeq-12$ to $-2\Delta\ln\mathcal{L}\simeq-6$ relative to $\Lambda$CDM, making the evidence for this model marginal. Since most of the improvement to the total likelihood comes from considering data sets (Planck CMB vs.~Planck SZ) that are already in tension within the $\Lambda$CDM paradigm, it is not surprising that adding another data set which is not in tension with $\Lambda$CDM decreases the significance. The fact that the Lyman-$\alpha$ data shows no preference for matter power spectrum suppression while at the same time being fully consistent with the Planck CMB data indicates that the nonvanishing value of $\Gamma_0$ favored by the Planck SZ data might be caused by systematics. We caution, however, that further analyses are necessary to reach a definitive conclusion, especially since other probes show a slight preference for a low value of $\sigma_8$ (see e.g.~Ref.~\cite{2017MNRAS.467.3024L}).

The main reason why the LSS data used in this analysis (following Ref.~\cite{Lesgourgues:2015wza} for direct comparison) strongly favor the presence of nonvanishing DM-DR interactions is that the information from Planck SZ and CHFTLenS are introduced as a direct Gaussian prior on $\sigma_8$. This direct constraint on what is essentially a derived cosmological parameter exacerbates the need for new DM physics. The ultimate analysis would be, instead, to go back to the actual measurement (e.g.~SZ cluster mass function or weak lensing shear correlation function) and perform the analysis directly in that space using the DM-DR interaction model. We leave such analysis to future work. In addition, we advocate for the need of verifying and testing systematics in each of these data sets before reaching any definite conclusion about the need of new physics. 

Since current data sets do not have enough sensitivity to confirm or rule out the presence of DM-DR interaction, we have performed a Fisher forecasts to test the constraining power of upcoming observations on the DM-DR interactions model. Using galaxy clustering measurements from the LSST photometric survey and CMB measurements from Stage-IV experiments, we find that constraints on the parameter $10^7 \Gamma_0$ should improve by a factor of $~\sim6$ compared to constraints from current data.  Also, our analysis show that the constraints on $\Delta N_{\rm fluid}$ could be improved by an order of magnitude compared to current constraints. 

In this work, we have used the Lyman-$\alpha$ measurements from Refs.~\cite{McDonald:2004eu,McDonald:2004xn} which constrains the amplitude of the matter power spectrum around $k\sim1h/$Mpc. While more recent Lyman-$\alpha$ forest measurements exist (see e.g.~\cite{2017PhRvD..96b3522I}), the absence of a likelihood code to compare the interacting DM-DR model predictions with these data makes including them in our analysis difficult. Given the tight constraints that these newer Lyman-$\alpha$ data put on warm DM models, it is possible that they could further constrain, and even rule out, the type of interacting DM-DR model considered in this work. 
As a rough guide, assuming that the inferred mean value of the matter power spectrum at $k\sim1h/$Mpc from the Lyman-$\alpha$ measurement stays the same, we estimate that a $\sim$60\% reduction of the error bar could exclude the DM-DR interaction model at the 3-$\sigma$ level.

\acknowledgments{We thank Azadeh Moradinezhad Dizgah for discussions. F.-Y. C.-R.~acknowledges the support of the National Aeronautical and Space Administration ATP grant NNX16AI12G at Harvard University.}

\bibliographystyle{JHEP}
\bibliography{dmdr_ref}

\end{document}